\documentclass{JHEP3} 
\usepackage{epsfig}
\title {$G_2$ gauge theory at finite temperature}
\author{Guido Cossu\\
        Scuola Normale Superiore, Piazza dei Cavalieri 27, 56126 Pisa\\
        and INFN Pisa, Largo B. Contecorvo 3 Ed.~C, 56127 Pisa, Italy\\
        E-mail: \email{g.cossu@sns.it}}
\author{Massimo D'Elia\\
        Dipartimento di Fisica \& INFN Genova, Via Dodecaneso 33, 16146 Genova, Italy\\ 
        E-mail: \email{delia@ge.infn.it}}
\author{Adriano Di Giacomo\\
        Dipartimento di Fisica \& INFN Pisa, Largo B. Contecorvo 3 Ed.~C, 56127 Pisa, Italy\\
        E-mail: \email{adriano.digiacomo@df.unipi.it}}
\author{Biagio Lucini\\
        Department of Physics, Swansea University, Singleton Park, Swansea SA2 8PP, UK\\
        E-mail: \email{b.lucini@swansea.ac.uk}}
\author{Claudio Pica\\
        Physics Department, Brookhaven National Laboratory, Upton, NY 11973, USA\\
        E-mail: \email{pica@bnl.gov}}
\abstract
{The gauge group being centreless, $G_2$ gauge theory is a good laboratory for studying the role of the centre of the group for colour confinement in Yang-Mills gauge theories. In this paper, we investigate $G_2$ pure gauge theory at finite temperature on the lattice. By studying the finite size scaling of the plaquette, the Polyakov loop and their susceptibilities, we show that a deconfinement phase transition takes place. The analysis of the pseudocritical exponents give strong evidence of the deconfinement transition being first order. Implications of our findings for scenarios of colour confinement are discussed.}
\keywords{Lattice Gauge Field Theories, Confinement}
\preprint{IFUP-TH/2007-22, BNL-NT-07/36}
\begin{document}
\section{Introduction}
Confinement is one of the most elusive problems in QCD. There is strong experimental evidence that quarks and gluons, which are the fundamental degrees of freedom of the theory, never appear as final states of strong interactions. It is still a challenge to understand how confinement is encoded in the QCD Lagrangian.\\
Following the large number of colours idea~\cite{'tHooft:1973jz}, it is reasonable to conjecture that confinement is a property of the gauge sector of the theory. Hence, it should be possible to solve the problem by looking at the pure gauge theory, and the solution should not be specific to a given number of colours $N$. For the pure gauge theory at finite temperature, it has been shown that confinement is lost at some critical temperature $T_c$~\cite{Borgs:1983wb}. The deconfinement phase transition in SU($N$) gauge theories can be understood in terms of the centre of the gauge group, which is $\mathbb{Z}_N$. An order parameter for the phase transition is the Polyakov loop
\begin{equation}
L(\vec{x},T) = \frac{1}{n} \mbox{Tr} \exp\left( i g \int_0^{1/T} A_0 \mbox{d} t \right) \ ,
\end{equation}
where $A_0$ is the gauge field in the compact direction, naturally associated
to the temperature $T$, whose length is $1/T$, $g$ is the gauge coupling and
$n$ the dimension of the fundamental representation (in SU($N$), $n =
N$). Since one dimension is compact, gauge transformations which are
continuous modulo $2 \pi/g$ are acceptable in the theory. Under those
transformations, $L(\vec{x},T) \to z L(\vec{x},T)$, where $z$ is an element of
$\mathbb{Z}_N$. If the centre symmetry is not broken, $\langle L \rangle =
(1/V)\int L(\vec{x},T) d^3 x = 0$ in the thermodynamic limit $V \to \infty$,
$V$ being the volume of the system. Conversely, a value of $\langle L \rangle$
different from zero implies breaking of the centre symmetry. It is possible to
show that at low temperatures $\langle L \rangle = 0$, while at high
temperatures $\langle L \rangle \ne 0$. Hence, a centre symmetry breaking phase
transition must take place. The expectation value of the Polyakov loop can be
related to the free energy $F$ of a static quark as
\begin{eqnarray}
L \propto e^{- \beta F} \ .
\end{eqnarray} 
It is then natural to identify the centre symmetry breaking phase transition with the deconfinement phase transition. In a famous paper~\cite{Svetitsky:1982gs}, Svetitsky  and Yaffe conjectured that the universality class of the deconfinement phase transition for SU($N$) gauge theory in D=d+1 dimensions is that of a d-dimensional $\mathbb{Z}_N$ Potts model, provided that the latter has a second order phase transition.  The Svetinsky-Yaffe conjecture has been verified numerically in 3+1 and 2+1 dimensions (see~\cite{Lucini:2003zr,Liddle:2005qb} for recent lattice calculations). It is interesting to remark that whenever the underlying spin model has a first order phase transition, so does the SU($N$) gauge theory.\\
This analysis hints toward the relevance of the centre for confinement. An
independent way to relate centre symmetry and confinement is presented
in~\cite{'tHooft:1977hy}, where confinement is described in terms of
condensation of vortices carrying magnetic flux. The allowed $N$ magnetic
fluxes are in one to one correspondence with the centre elements of the
group. Condensation of vortices in the confined phase means that the area
spanned by a Wilson loop randomly intersect vortex worldsheets. The resulting
cancelations determine the so-called area law for the Wilson loop, which is
one of the accepted criteria for colour confinement. Numerical works have
confirmed the vortex scenario~\cite{DelDebbio:1996mh}. To characterise the
transition in terms of a symmetry, the 't Hooft loop operator can be
introduced~\cite{'tHooft:1977hy}, which is expected to have a non-zero vacuum
expectation value in the confined phase and to be zero in average in the
deconfined phase. This behaviour has been checked numerically
in~\cite{deForcrand:2000fi,DelDebbio:2000cx,DelDebbio:2000cb, Kovacs:2000sy}.\\
While this scenario for colour confinement is perfectly consistent, the centre symmetry is lost when dynamical fermions are added to the action. Hence, either one gives up the idea that confinement in the pure Yang-Mills theory and in the full theory is basically the same phenomenon or we must assume that the centre is just a useful way to look at confinement, but does not embody any fundamental physics in relation to it. One possible way to look at this issue is to study the deconfinement phase transition in other gauge groups that have a different centre pattern. The physics of the phenomenon being inherently non-perturbative, lattice calculations are well suited for those investigations. In this context, SO($3$)$\equiv$ SU(2)/$\mathbb{Z}_2$ would be an ideal candidate: it is expected to confine (like SU(2), since the two groups share the same algebra), but has a trivial centre. Recent results suggest that a deconfinement phase transition takes place, but the presence of lattice artifacts ({\em centre monopoles}) makes it difficult to extract a reliable continuum limit~\cite{Barresi:2003jq}. Moreover, the centre structure of the underlying universal covering group (SU(2)) reflects in the existence of twist sectors, which might imply that the centre still plays a role, despite the group being centreless.\\
A different way to approach the problem is to use a fundamental group that is genuinely centreless\footnote{We use the word centreless to refer to a group whose centre is given only by the unity element.}. The simplest group in this category is the exceptional group $G_2$. There are other properties that make $G_2$ interesting for QCD: it contains SU(3) as a subgroup and (as in full QCD) an asymptotic string tension does not exist, since the colour charge carried by a quark can be completely screened by gluons~\cite{Holland:2003jy}. The existence of two phases has been proved in~\cite{Holland:2003jy}. However, this does not exclude that, instead of a real phase transition, a crossover separates the two phases. Were this the case, the physics of deconfinement in $G_2$ would be noticeably different from that of SU($N$) gauge theories, and this would cast serious doubts about what we can learn from $G_2$ for confinement in more physical gauge theories. While data reported in~\cite{Pepe:2005sz,Pepe:2006er} are compatible with a first order phase transition taking place, no exhaustive and detailed study of deconfinement has been performed so far. In this paper, we shall fill this gap by studying the finite size scaling behaviour of the plaquette, of the Polyakov loop and of their susceptibilities, from which we extract the critical exponents for the transition. We will then be able to show that a real transition takes place and that this transition is first order.\\
This work is organised as follows. In Sect.~\ref{g2} we will review the basic properties of the exceptional group $G_2$. Details of our lattice simulations are presented in Sect.~\ref{details}. Sect.~\ref{thermodynamics} contains our results and provides evidence for a first order deconfinement phase transitions occurring in $G_2$ at finite temperature. The implications of our findings for possible mechanisms of colour confinement are discussed in Sect.~\ref{discussions}. Finally, in Sect.~\ref{conclusions} we summarise the main points of our investigation. 
\section{Basic properties of the exceptional group $G_2$}
\label{g2}
We begin by summarising some basic properties of the Lie Group $G_2$. In mathematical terms this is the group of automorphisms of the octonions and it can be naturally constructed as a subgroup of the real group $SO(7)$ - which has 21 generators and rank 3. Besides the usual properties of $SO(7)$ matrices
\begin{equation}
  \det \Omega = 1 \qquad \Omega^{-1} = \Omega^{T}
\end{equation}
we have in addition another constraint
\begin{equation}
  T_{abc} = T_{def} \Omega_{da} \Omega_{eb} \Omega_{fc}
  \label{constraint}
\end{equation}
where $T_{abc}$ is a totally antisymmetric tensor whose nonzero elements are (using the octonion basis given by \cite{Cacciatori:2005yb})
\begin{equation}
  T_{123} = T_{176} = T_{145} = T_{257} = T_{246} = T_{347} = T_{365} = 1.
\end{equation}
Equations~(\ref{constraint}) are 7 independent relations reducing the numbers of generators to 14. The fundamental representation of $G_2$ is 7 dimensional. Using the algebra representation of \cite{Cacciatori:2005yb} (we refer to appendix \ref{G2algebra} for details) we can clearly identify an $SU(3)$ subgroup and several $SU(2)$ subgroups,  6 of which are sufficient to cover the whole group, a useful property for MC simulations. The first three $SU(2)$ subgroups are in the $4 \times 4$ real representation of the group while the remaining three are a mixture of the $4 \times 4$ and the $3 \times 3$ representations and are extremely difficult to simulate with standard heat-bath techniques. See the next section for details on simulations.\\
The following relations hold:
\begin{equation}
  SU(3) \subset G_2  \Rightarrow \mathcal C(G_2) \subset \mathcal {\rm{Centr}}(SU(3)) = \mathbb Z_3
\end{equation}
in which ${\rm{Centr}}(SU(3))$ is the centralizer of $SU(3)$ (i.e. the
matrices in $G_2$ that commute with every element in $SU(3)$). Intersections of
centralizers of different $SU(3)$ subgroups give
\begin{equation}
 \mathcal C(G_2)= \{1\}
\end{equation}
 i.e. a trivial centre.\\
The Lie group $G_2$ has rank 2, like $SU(3)$. This implies that the residual
symmetry after an Abelian projection is $U(1)^2$, its Cartan subgroup. Stable monopole solutions are classified according to the homotopy group\footnote{The first equality follows from $\pi_1(G_2) = 0$. See for example~\cite{Weinberg:1996kr}.}:
\begin{equation}
  \pi_2(G_2/U(1)^2) = \pi_1(U(1) \times U(1)) = \mathbb Z \times \mathbb Z 
\end{equation}
i.e. we have two distinct species of monopoles, classified by elements of the discrete group $\mathbb Z_2$, as for $SU(3)$. 
An extension of the 't Hooft tensor - the gauge invariant field of monopoles - can be written for the $G_2$ gauge group so Abelian monopole solutions are really possible in this theory.\\
Another interesting homotopy group shows that centre vortices are absent in the theory:
\begin{equation}
  \pi_1(G_2/\mathcal C(G_2)) = \pi_1(G_2) = 0
\end{equation}
while for $SU(3)$ for example
\begin{equation}
  \pi_1(SU(3)/\mathbb Z_3) = \mathbb Z_3
\end{equation}
and 
\begin{equation}
  \pi_1(SO(3)/\{ 1\}) = \pi_1(SO(3)) = \mathbb Z_2 \neq 0
\end{equation}
as stated before. So $G_2$ is a good playground to study the dual superconductor picture in a theory without centre vortices, thus isolating monopole contribution in confinement.
\section{Simulations of $G_2$ Lattice Gauge Theory}
\label{details}
In this work we are going to investigate the thermodynamical properties of the gauge group $G_2$ (see also \cite{Holland:2003jy, Pepe:2005sz}). To simulate the pure gauge theory
\begin{equation}
\mathcal L = \frac{1}{7g^2} {\rm Tr} \, F_{\mu\nu}F_{\mu\nu}
\end{equation}
with the Wilson action, we used a simple Cabibbo-Marinari update (heat-bath + overrelaxation in a tunable ratio, for every step) for the first three $SU(2)$ subgroups ($4 \times 4$ representation, set 1,3 and 4 in appendix A) spanning the $SU(3) \subset G_2$. This simple setting cannot be used for the remaining three subgroups because the integration measure is not as simple. We make a random gauge transformation every $n$ updates (tipically 1 or 2) to guarantee the ergodicity of the algorithm\footnote{The matrices for random gauge transformation are regenerated every step by a random algorithm to assure that no periodicities or orbits in phase space can arise.}.
To study the thermodynamical properties we simulated several
asymmetric lattices $N_t \times N_s^3$ of spatial dimensions $N_s =
12, 14, 16, 18, 20, 24, 32$ and temporal dimension $N_t = 6$ ($N_t =
4$ only for the smallest lattice). An average of 20 $\beta$s per lattice have
been simulated. The temperature of the system is
given by $T = (a(\beta) N_t)^{-1}$, where $a(\beta)$ is the lattice spacing
as a function of $\beta = 1/7 g^2$. The critical behaviour of the
system has been extracted by applying the theory of finite size
scaling (FSS), which has been used to extrapolate the behaviour of the
observables we have studied to the thermodynamic limit ($N_s \to \infty$). We
needed histories of order $10^5$ updates near the transition (1 week on a
1.5GHz Opteron processor for a medium lattice).\\
The code is highly optimized and very fast (using only real algebra), is
written using explicitly assembler SSE2 instructions in single precision for
the matrix-multiplication core and run on an Opteron farm in the computer
facilities of the Physics Department of the University of Pisa. 

The observables we have measured are the standard plaquette and the Polyakov loop. A clarification is in order here. While one should expect to be able to characterise the critical behaviour of a system by looking at the plaquette, doubts could be cast into the usefulness of the Polyakov loop: since $G_2$ is centreless, the Polyakov loop is not an order parameter for a possible deconfining phase transition. In principle, phase transitions can be reliably investigated only by using an order parameter field, whose critical behaviour characterises the transition itself. However, in order to prove that a transition takes place and to determine the critical indices, a non-trivial overlap on the order parameter is the only property we need\footnote{The reverse of this sentence is not true: no conclus§ion can be drawn from the absense of critical behaviour in a non-order parameter field.}. Hence, if we can observe a divergence in the peak of the Polyakov loop susceptibility (and of the specific heat, whose reliability is hard to question) we can safely conclude that a phase transition takes place.\\  
The theory of FSS predicts that as a function of the volume the
maximum of susceptibilities scale in the following way:
\begin{equation}
\label{fss}
\chi \sim a \cdot L^{\frac{\gamma}{\nu}} + b \ ,
\end{equation}
where $\gamma$ is the critical exponent of the generating quantity (in
our case either the plaquette or the Polyakov loop) and $\nu$ is the
critical exponent related to the divergence of the correlation
length. 
The position of the maximum scales as
\begin{eqnarray}
\label{betac}
\beta_c(L) = \beta_c(\infty) + c L^{-1/\nu} \ ,
\end{eqnarray}
where $\beta_c(L)$ is the pseudocritical $\beta$ for size $L$ and
$\beta_c(\infty)$ is the critical value of $\beta$.
This analysis also applies to first order phase transition,
whose signature is given by $\gamma = 1$ and $\nu = 1/d$, with $d$ the
dimension of the system. \\

\section{Thermodynamics of $G_2$ gauge theory}
\label{thermodynamics}
We studied the thermodynamics of this theory using the typical observables, the plaquettes 
\begin{equation} 
P_s = \frac{1}{3 \cdot 7 N_s^3 N_t} \sum_{\square_s} {\rm Tr} U_{\square_s} \qquad  P_t = \frac{1}{3 \cdot 7 N_s^3 N_t} \sum_{\square_t} {\rm Tr} U_{\square_t}
\end{equation}
where the two sums are  on space-space and space-time plaquettes respectively. The peak of the susceptibility
\begin{equation}
\chi_P = N_s^3 (\langle P^2 \rangle - \langle P \rangle^2) \qquad P = (P_s + P_t)/2
\end{equation}
signals the phase transition point. This quantity (often referred to in the literature as the "lattice specific heat") is only part of the (physical) specific heat, whose complete reconstruction requires various correlators weighted with different coefficients; nonetheless, this is a singular piece from which the critical scaling behaviour can be inferred.


We also measured the Polyakov loop and its susceptibility:
\begin{equation}
L = \frac {1}{N_s^3}\sum_{\vec x}\Bigl (\frac {1}{7}\prod_{t=0}^{N_t-1} U_4 (\vec x) \Bigr ) \qquad \chi_L = N_s^3 (\langle L^2 \rangle - \langle L \rangle^2).
\end{equation}
\FIGURE[ht]{
  \epsfig{file=./FIGS/Plaquette_susceptibility.eps,width=0.85\linewidth, clip=}
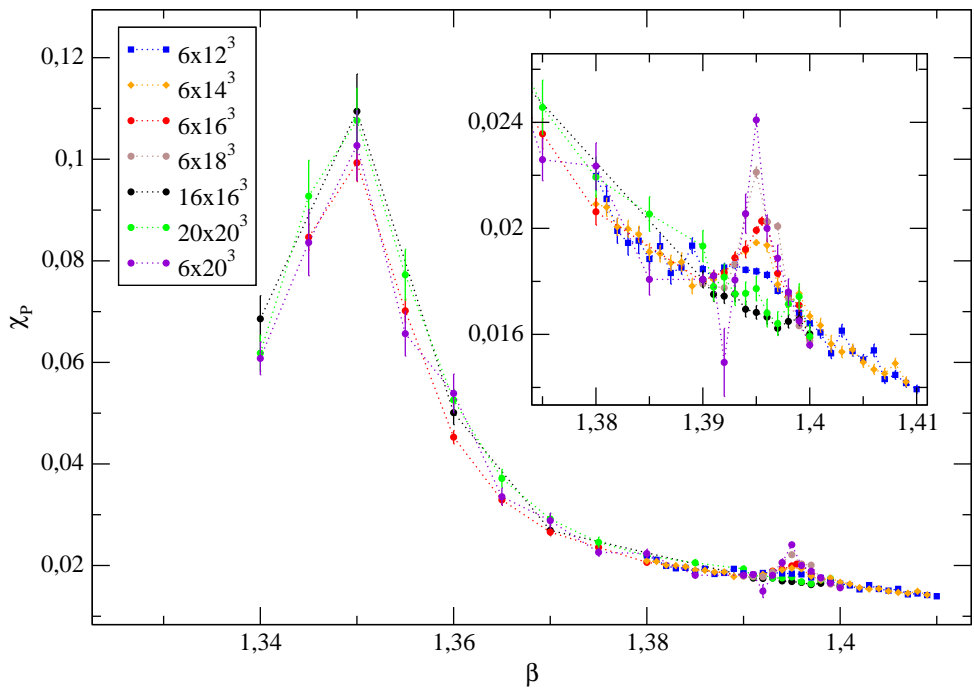
  \caption{Plaquette susceptibility plotted against $\beta$. The peak signals
  the bulk transition while the peak corresponding to the physical transition
  for $N_t = 6$ is shown in the inset. We also show results from a simulation at T=0 on a $16^4$ lattice (black triangle points).}
  \label{susc-plaq}
}
\FIGURE[ht]{
  \epsfig{file=./FIGS/Plaquette_bulk_transition.eps,width=0.85\linewidth,clip=}
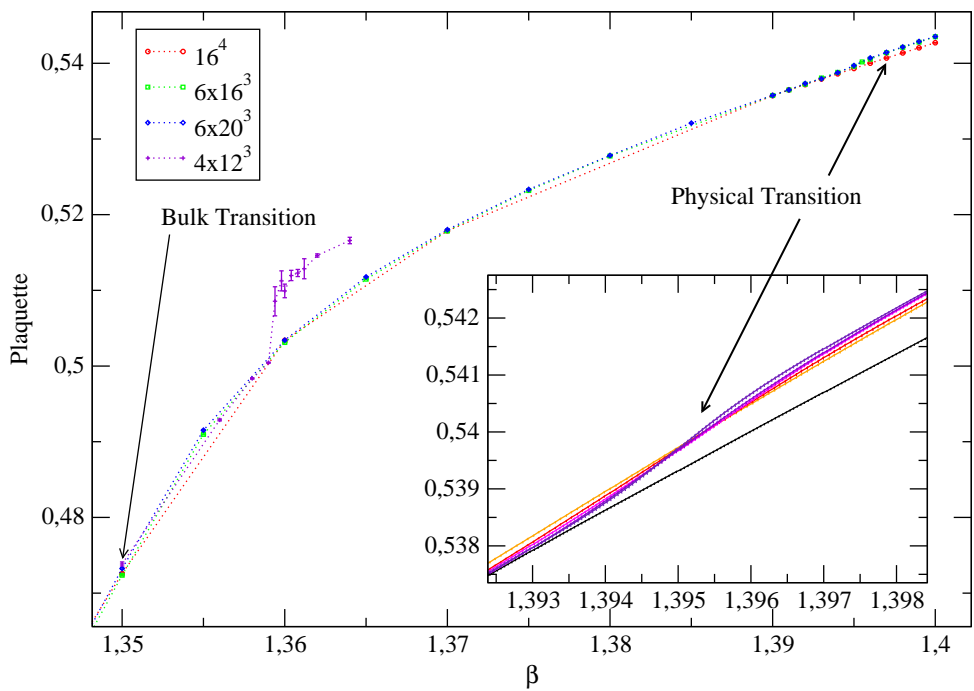
  \caption{Comparison of finite and zero temperature simulations. In the box:    magnification of the physical transition region (reweighted curves).}
  \label{Comparison}
}
\FIGURE[ht]{
\centering
\begin{tabular}{cc}
  \epsfig{file=./FIGS/Plaquette_Peak_Scaling.eps,width=0.45\linewidth,clip=}&
  \epsfig{file=./FIGS/Plaquette_susceptibility_reweighted_rescaled.eps,width=0.48\linewidth,clip=}
\end{tabular}
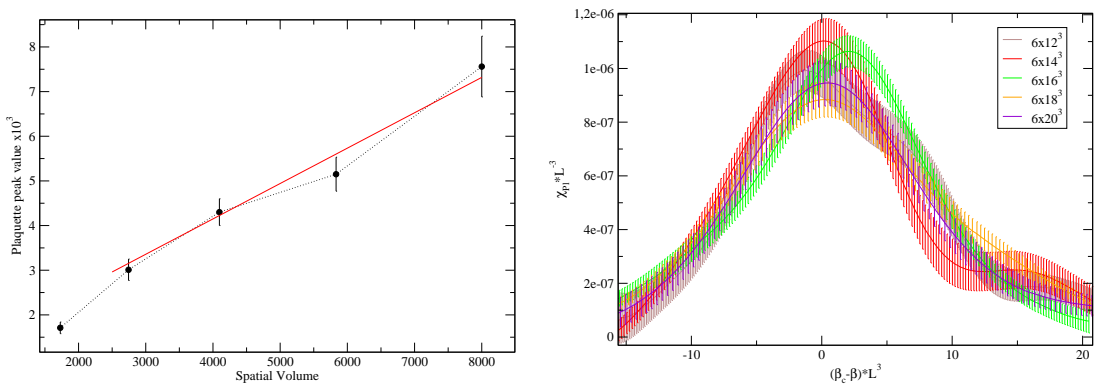
  \caption{Left: scaling of the peak of plaquette susceptibility with the
  volume. The continuous line is a linear fit to the data, as
  explained in the text. Right: FSS of the plaquette susceptibility assuming a
  first order transition. For this plot, we have used the value $\beta_c = 1.395$, obtained from the fit to the position of the maximum according to~(\ref{betac}).} 
    \label{Plaq_scaling}
}
\FIGURE[ht]{
  \epsfig{file=./FIGS/Polyakov_Loop_Density_12.eps,width=0.85\linewidth, clip=}
  \vspace{20pt}
  \\
  \epsfig{file=./FIGS/Polyakov_Loop_Density_16.eps,width=0.85\linewidth, clip=}
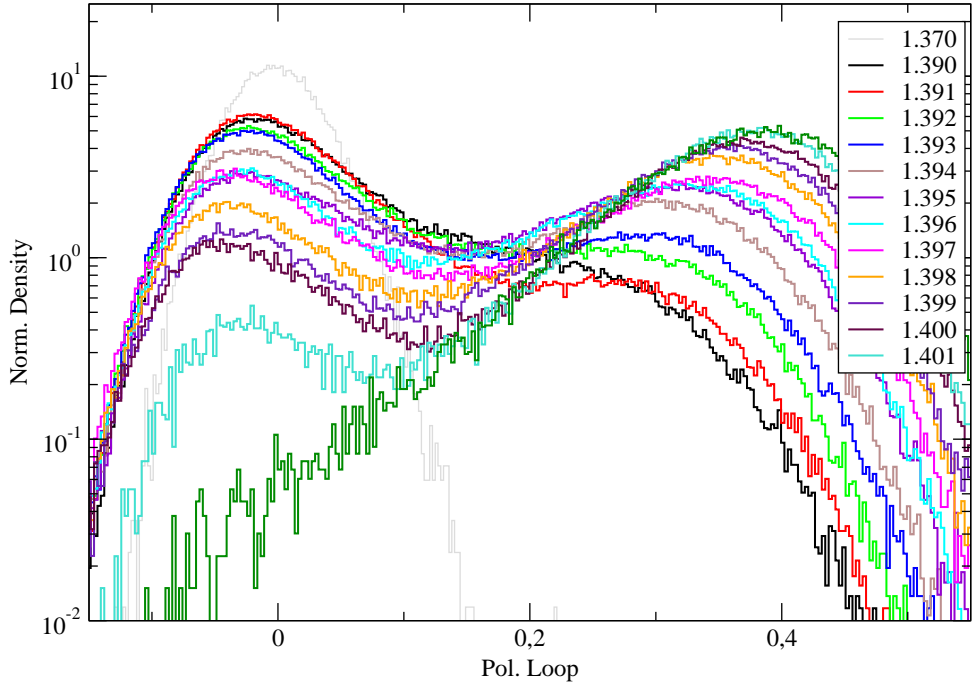
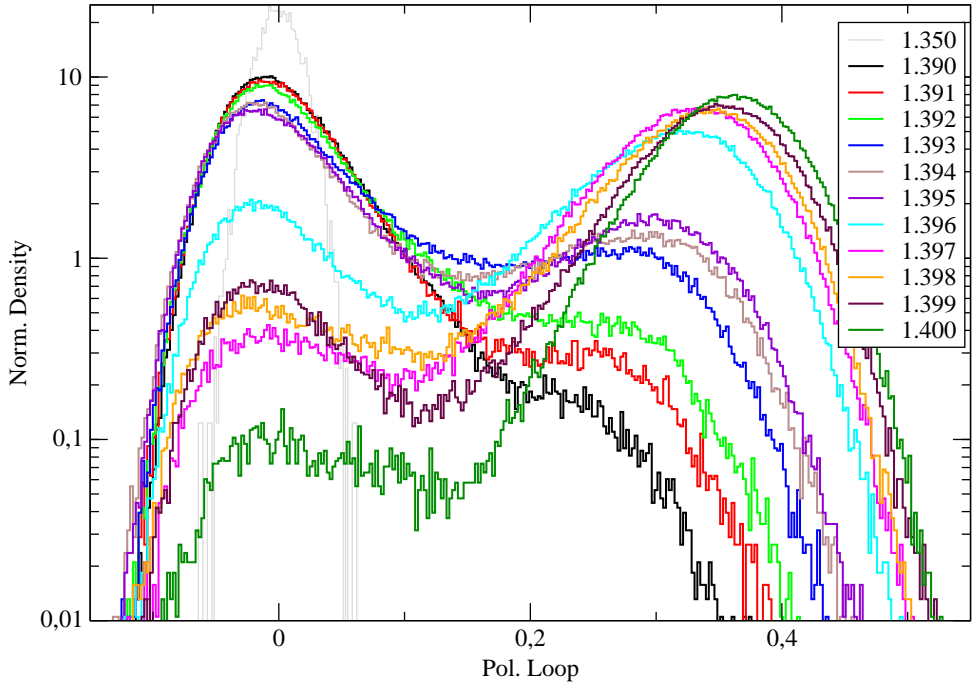
  \caption{Normalized densities of the Polyakov Loop in a semilog plot for
    $\beta$ varying in the range from 1.35, the critical coupling of the bulk
    transition ``$\beta_{bulk}$'', to 1.401, in the deconfined phase (data
    from the $6\times14^3$ lattice for the upper graph and from $6\times16^3$
    for the other - same scales and limits for both axes are used for better
    comparison). As an aside we notice that far in the confined phase,
    $\beta_c < 1.395$, the Polyakov loop is zero within errors and this feature can not be explained on the ground of any manifest symmetry of the system. Continues on next page}
  \label{PolDensity1}
}
\FIGURE[th]{
  \epsfig{file=./FIGS/Polyakov_Loop_Density_20.eps,width=0.85\linewidth, clip=}
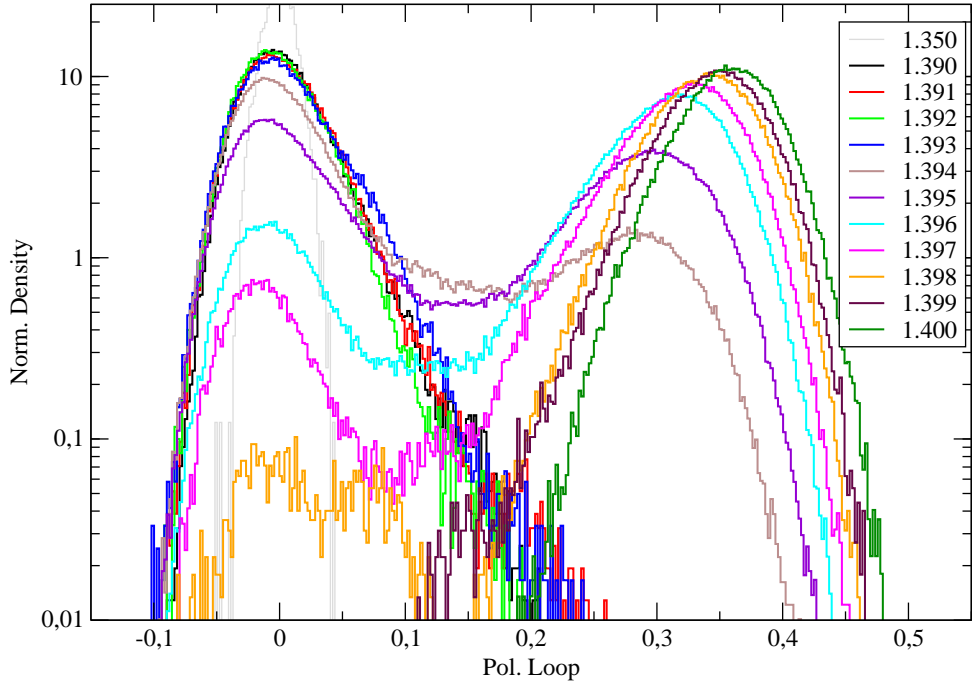
  \caption{Continues from last page ($6\times20^3$ lattice).}
  \label{PolDensity2}
}
The lattices considered for the scaling analysis are only the $N_s = 12, 14, 16, 18,
20$ times $N_t = 6$ for the following reasons. The computational cost of
locating the transition grows exponentially fast with the volume; anticipating here a first
order transition, the intrinsic problem is that two (or more) phases
coexist. The simulated system tunnels between pure phases by building an
interface of size $N_s$. The free-energy cost of such a mixed configuration is
$\sigma N_s^{D-1}$ ($\sigma$ being the surface tension), the interface is
built with probability $\exp(-\sigma N_s^{D-1})$ and the natural time scale
for the simulation grows with $N_s$ as $\exp(\sigma N_s^{D-1})$. This is
called exponential critical slowing down and makes simulations impractical for
lattices with $N_s > 20$ for a reliable estimate of susceptibilities. Looking
at Figs. \ref{PolDensity1}, \ref{PolDensity2} and comparing the densities in the tunneling region
for the three different lattices gives an idea of the problem, common to all
systems exhibiting a first order transition. Multicanonical methods
\cite{Janke:1998} will be needed for feasible simulations on such large
lattices. The other reason concerns the number of time slices and is related
to the presence of an unphysical bulk transition that we shall explain below (see also Fig. \ref{susc-plaq}). Being very close to the bulk transition, the
physical deconfinement transition for $N_t = 4$ is extremely difficult to
detect, the signal being highly contaminated by the ``noise'' coming from the
bulk. $N_t = 6$ is needed to be sufficiently away from the bulk. By increasing furtherly $N_t$, one can move
the physical transition far away from the bulk transition point. Hence,
choosing a larger $N_t$ will clean the signal from the bulk ``noise''. To
investigate this possibility, we performed some simulations at $N_t = 8$,
which confirmed the general features of the $N_t = 6$ simulation. The
displacement of the critical $\beta$ was clearly visible but not sufficient to
bring any practical advantage over the $N_t = 6$ calculation, while the simulation time increased considerably. For this reason, we sticked to the $N_t = 6$ calculation, giving up the possibility of performing a continuous limit extrapolation of the critical temperature. However, our pilot study at $N_t = 8$ suggests that there is no reason to doubt that such a continuous limit exists.\\
\FIGURE[ht]{ 
  \centering
  \begin{tabular}{cc}
    \epsfig{file=./FIGS/Plaquette_History_4x12_1.3594.eps,width=0.45\linewidth,
      clip=}&
    \epsfig{file=./FIGS/Polyakov_Loop_History_6x20_1.395.eps,width=0.45\linewidth,clip=}
  \end{tabular}
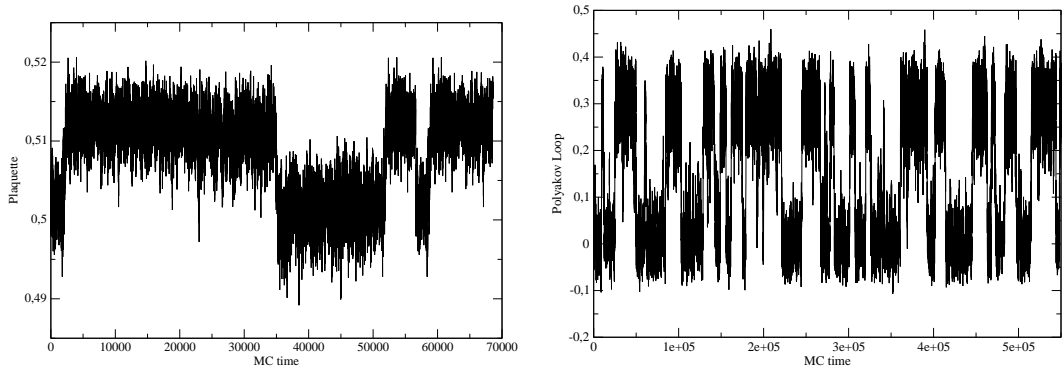
  \caption{Left: MC history of the plaquette ($\beta = 1.3594,
      12^3\times4$). Right: A typical Monte Carlo history of the Polyakov loop (data from
      $\beta = 1.395, 20^3\times6$). }
  \label{Histories}
}
\FIGURE[th]{
  \epsfig{file=./FIGS/Polyakov_Loop_Rescaling.eps,width=0.85\linewidth, clip=}
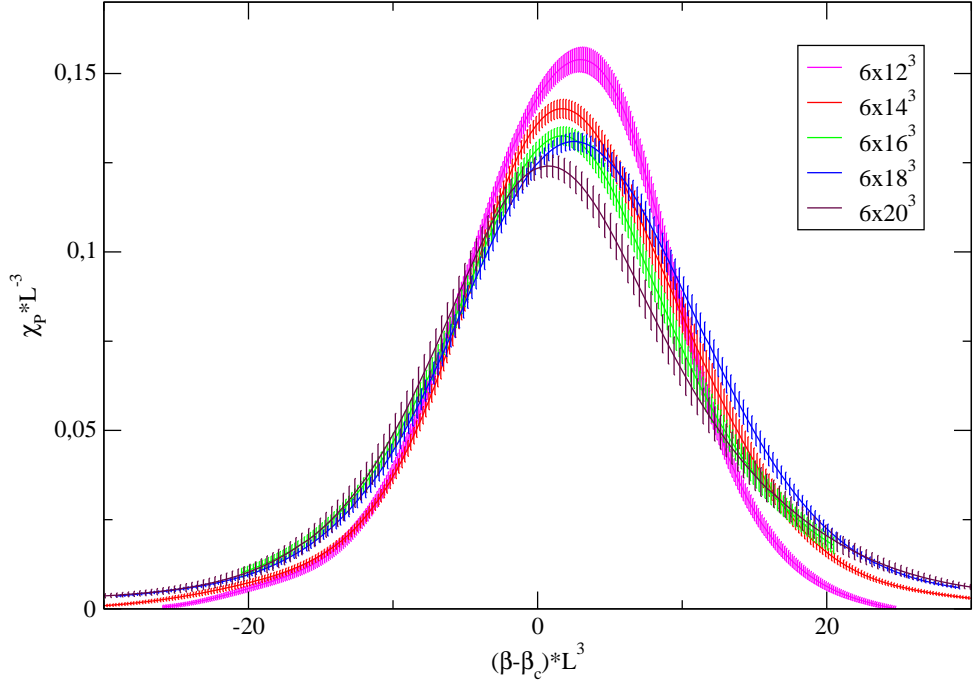
  \caption{Scaling of the Polyakov loop assuming first order. For the smallest
    lattice $12^3\times 6$ corrections to the scaling are evident (even the
    lattice $14^3\times 6$ is not big enough but corrections are reduced); $\beta_c = 1.395$ as explained in the text.}
\label{RescalingPol_up}
}
\FIGURE[b]{
  \epsfig{file=./FIGS/Polyakov_Loop_Peak_Scaling.eps,width=0.85\linewidth, clip=}
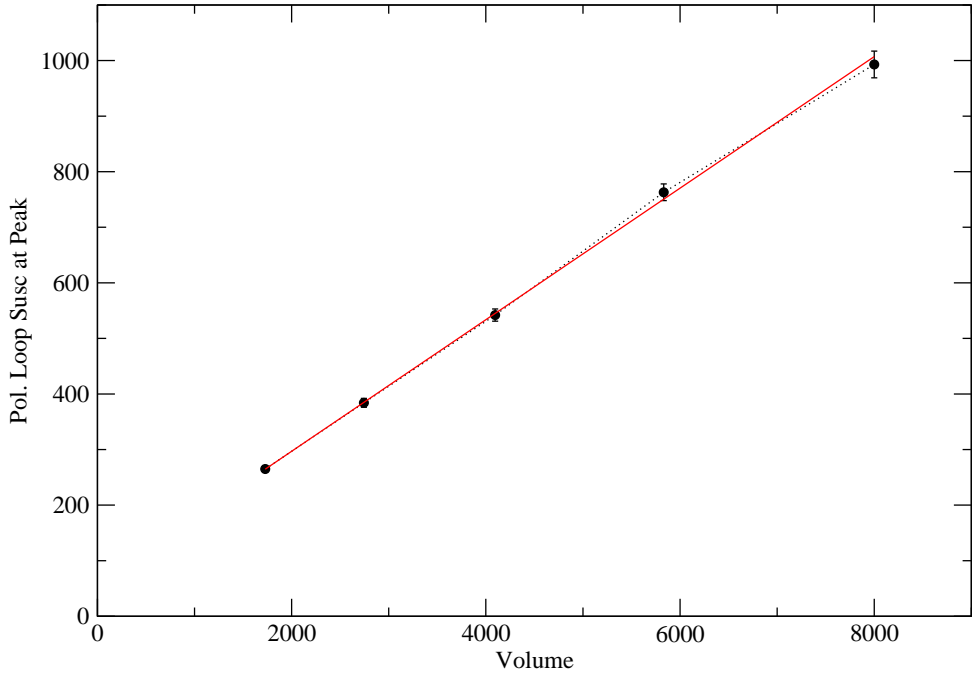
  \caption{Scaling of the peak of $\chi_L$. The solid line is a linear
  fit to the data.} 
  \label{RescalingPol_down}
}
In a finite volume no divergences can arise, since the partition function is analytical. Nevertheless critical indices can be measured by looking at the scaling with the volume of the plaquette susceptibility (related to the specific heat $C_V$). The height of the peak for a first-order transition scales with the volume $V$ and the width and the displacement from the real critical point of the peak position scales as $1/V$ (plus corrections to this leading behaviour).\\
A pronounced peak is present at any volume and $N_t$ and always at the same
$\beta \sim 1.35$. There is no scaling with volume and no movement toward the
weak coupling region passing from $N_t = 4$ to $N_t = 6$ as we would expect
for a physical transition. This transition is the equivalent of the bulk phase
transition in SU($N$) gauge theories, and separates the (physical) weak
coupling region from the (unphysical) strong coupling one. The bulk peak
almost completely overshadows the real physical transition, a smaller peak in
the weak coupling region at $\beta \sim 1.395$ for $N_t = 6$. This peak scales
with the volume, provided that the bulk contribution has been subtracted. This
subtraction procedure is needed in order to disentangle the physics from the
discretisation artifacts. To estimate the bulk background, we simulated the
system also at zero temperature on $16^4$ and $20^4$ lattices (to control
systematic errors). The bulk contribution has
to be subtracted from the plaquette susceptibility for a correct finite
scaling analysis. This procedure could be seen as a normalisation of the free
energy following the request that this quantity be zero at zero
temperature. The influence of the bulk transition on the plaquette
susceptibility is shown in Fig. \ref{susc-plaq}. The nature of the two transitions manifests itself comparing finite temperature and zero temperature simulations in Fig. \ref{Comparison}. The integral of the difference between the two curves is the free energy density:
\begin{equation}
  \frac{f}{T^4}\Bigr|_{\beta_0}^{\beta} = - N_\tau^4 \int_{\beta_0}^\beta d\beta^\prime (P_0 - P_T)
\end{equation}
in which $P_0$ and $P_T$ are the mean plaquettes at zero and finite temperature
respectively.  At the bulk transition $f$ is zero within errors and
develops a value different from zero at the physical transition.\\
The MC time history of the plaquette is
displayed in Fig.~\ref{Histories} (left), and shows a two-phase
structure typical of first order phase transitions.
The extracted maxima of the plaquette susceptibility ($\propto C_V$) using the reweighted data are shown in
Fig. \ref{Plaq_scaling}. Maxima and their errors are estimated by a simple
inspection of the reweighting output. A linear fit of the form $y = a \cdot x + b$
(see Eq.~\ref{fss}) gives $a = 0.00079(14)\cdot10^{-3},\, b = 0.98(62)\cdot10^{-3},\,
\chi^2_{\rm red} = 1.35$, providing good evidence for a first order
phase transition. A fit according to Eq.~(\ref{betac}) gives $\beta_c(\infty) = 1.3950(4)$.\\
The Polyakov loop is insensitive to the bulk transition so we used it
to detect the position of the physical one, even if, strictly
speaking, this quantity is not an order parameter. The Polyakov loop develops
an evident double peak structure typical of a first order transition (see
Figs.~\ref{PolDensity1},\ref{PolDensity2}). In this semilog plot is also clear, by looking
at the relative ratio of peaks height and valley height near the
transition point, the exponential decreasing of tunneling
probability with the volume. In Fig.~\ref{Histories} we show the typical
Monte Carlo history of the Polyakov loop. Once again, a clean two-state
signal appears. This reflects in a double-peak structure of the
observable shown e.g. in Fig.~\ref{PolDensity2}. The same FSS analysis as for the specific heat
again gives evidence of a first order transition, with a good $\chi_{\rm red}^2$ in the
linear fits of peak heights (Figs.~\ref{RescalingPol_up}~and~\ref{RescalingPol_down}). The parameters of the linear fit of the peak
heights $y = a \cdot x + b$ are $a = 0.1183(2),\, b = 60(5),\,
\chi^2_{\rm red} = 0.61$. A subtraction of the background is
understood. The background is assumed to be weakly dependent on coupling
$\beta$. This is an educated guess suggested by the zero temperature
simulations. The background is estimated by mean of a linear fit of the tails
of the peak and being an ultraviolet effect, it is assumed to be the same for
all volumes. In practice we took the smallest lattices $6 \times 12^3$, $6
\times 14^3$ and some
of the extremal points in tails for the fit. The number of points is
unessential giving practically the same parameters and a good $\chi_{\rm red}^2$. The Polyakov loop susceptibility can be also used to determine $\beta_c(\infty)$. Using formula~(\ref{betac}), we get $\beta_c(\infty) = 1.3951(2)$, which is compatible with the result obtained from the susceptibility of the plaquette.
\section{Discussion}
\label{discussions}
As we have stated in the introduction, an asymptotic string tension in
$G_2$ does not exist. Hence, one can question whether this group is
confining. This is mostly a semantic
problem. In~\cite{Greensite:2006sm} it is argued that because of the
absence of the asymptotic string, $G_2$ gauge theory is not
confining. This would fit the idea of confinement as related to centre
vortices randomly piercing the Wilson loop. Sharing this view means to
accept the logical conclusion that full QCD (in which an asymptotic
string tension does not exist because of quark pair production) is not
a confining theory. Since it is common understanding that QCD confines, the
essence of confinement must be found in some other property of the
theory. In our opinion, this property is a low-energy dynamics
dominated by glueballs and mesons (which are colour-singlet
states). Colour-singlet states are also present in $G_2$ at zero
temperature. At high temperature the dynamics is instead dominated by
a gluon plasma. In this sense, despite the absense of an asymptotic
string tension, $G_2$ gauge theory is a confining theory. Accepting this statement means to infer that centre degrees of freedom are not related to confinement (unless one want to put all the weight of the centre on the trivial element, see~\cite{Greensite:2006sm}). Hence, the degrees of freedom responsible for colour confinement must be searched for in other properties of the gauge group.\\
Like SU(3), $G_2$ is a rank two group, i.e. it has two Cartan generators\footnote{A Cartan generator is a generator which commutes with all the others.}. It is then an attractive possibility that like in SU($N$) pure gauge theories~\cite{DiGiacomo:1999fa,DiGiacomo:1999fb,Carmona:2001ja} and in full QCD~\cite{Carmona:2002ty,D'Elia:2005ta} the mechanism for colour confinement is related to the condensation of magnetic monopoles, as it seems to be the case also for the SO($3$) gauge theory~\cite{Barresi:2004qa}. An investigation in this direction is currently in progress, and will be reported elsewhere.
\section{Conclusions}
\label{conclusions}
We studied the thermodynamics of the Yang-Mills theory with gauge
group $G_2$. The presence of an unphysical transition (most probably
due to the choice of the discretised action used in simulations) makes
the problem harder. Nevertheless a physical transition is found by
looking at plaquette and Polyakov loop susceptibilities. Time
histories of the Polyakov group and the plaquette show double peaks
typical of first order transitions. A detailed FSS
analysis agrees with the first order hypothesis.
Hence, we can conclude that $G_2$ gauge theory has two distinct phases
separated by a jump in the free energy. Those phases are immediately
identified with the confined (low temperature) and deconfined (high
temperature) phase. The same dynamics characterises SU($N$) Yang-Mills
theories at finite temperature. Since $G_2$ does not have a
(non-trivial) centre, our findings suggest that the dynamics of colour
confinement cannot be directly related to the centre of the gauge
group, as it has been inferred from previous works on SU($N$) gauge
theories. At this stage, the possibility that dual superconductivity
of the vacuum explains colour confinement is still open. The next step
of our study is to investigate the FSS of the monopole creation
operator, to test if the dual superconductor picture of confinement
works also for $G_2$ gauge theory.

\acknowledgments
\label{acknowledgments}
We would thank M.~Pepe for various useful discussions on the topic. The work
of C.P. has been supported in part by contract DE-AC02-98CH1-886 with the
U.S. Department of Energy and B.L. is supported by the Royal Society.

\appendix
\section{$G_2$ algebra representation}
\label{G2algebra}
In this appendix we simply report a representation of the 14 generators of the $G_2$ group \cite{Cacciatori:2005yb}. They are normalized such that $\rm{tr} (C_i C_j) = - \delta_{ij}$. The first 8 matrices generate the $SU(3) \subset G_2$. Here is also a list of 6 $SU(2)$ subroups that cover the entire group (useful for the Cabibbo-Marinari update):
\begin{enumerate}
\item $C_1, C_2, C_3$
\item $\sqrt{3} C_8, \sqrt{3}C_9, \sqrt{3}C_{10}$
\item $C_4, C_5, \frac{(C_3 + \sqrt{3} C_8)}{2}$
\item $C_6, C_7, \frac{(C_3 - \sqrt{3} C_8)}{2}$
\item $\frac{(3 C_3 - \sqrt{3} C_8)}{2}, \sqrt{3}C_{11}, \sqrt{3} C_{12}$
\item $\frac{(3 C_3 + \sqrt{3} C_8)}{2}, \sqrt{3}C_{13}, \sqrt{3} C_{14}$
\end{enumerate}

\section{Algebra}
\small
$$
C_1 =\frac{1}{2}\left(
\begin{array}{ccccccc}
0 & 0 & 0 & 0 & 0 & 0 & 0 \\
0 & 0 & 0 & 0 & 0 & 0 & 0 \\
0 & 0 & 0 & 0 & 0 & 0 & 0 \\
0 & 0 & 0 & 0 & 0 & 0 & -1 \\
0 & 0 & 0 & 0 & 0 & -1 & 0 \\
0 & 0 & 0 & 0 & 1 & 0 & 0 \\
0 & 0 & 0 & 1 & 0 & 0 & 0
\end{array}
\right)
\qquad
 C_2 =\frac{1}{2}\left(
\begin{array}{ccccccc}
0 & 0 & 0 & 0 & 0 & 0 & 0 \\
0 & 0 & 0 & 0 & 0 & 0 & 0 \\
0 & 0 & 0 & 0 & 0 & 0 & 0 \\
0 & 0 & 0 & 0 & 0 & 1 & 0 \\
0 & 0 & 0 & 0 & 0 & 0 & -1 \\
0 & 0 & 0 & -1 & 0 & 0 & 0 \\
0 & 0 & 0 & 0 & 1 & 0 & 0
\end{array}
\right)
$$
$$
C_3 =\frac{1}{2}\left(
\begin{array}{ccccccc}
0 & 0 & 0 & 0 & 0 & 0 & 0 \\
0 & 0 & 0 & 0 & 0 & 0 & 0 \\
0 & 0 & 0 & 0 & 0 & 0 & 0 \\
0 & 0 & 0 & 0 & -1 & 0 & 0 \\
0 & 0 & 0 & 1 & 0 & 0 & 0 \\
0 & 0 & 0 & 0 & 0 & 0 & -1 \\
0 & 0 & 0 & 0 & 0 & 1 & 0
\end{array}
\right)
\qquad
C_4 =\frac{1}{2}\left(
\begin{array}{ccccccc}
0 & 0 & 0 & 0 & 0 & 0 & 0 \\
0 & 0 & 0 & 0 & 0 & 0 & 1 \\
0 & 0 & 0 & 0 & 0 & 1 & 0 \\
0 & 0 & 0 & 0 & 0 & 0 & 0 \\
0 & 0 & 0 & 0 & 0 & 0 & 0 \\
0 & 0 & -1 & 0 & 0 & 0 & 0 \\
0 & -1 & 0 & 0 & 0 & 0 & 0
\end{array}
\right)
$$
$$
C_5 =\frac{1}{2}\left(
\begin{array}{ccccccc}
0 & 0 & 0 & 0 & 0 & 0 & 0 \\
0 & 0 & 0 & 0 & 0 & -1 & 0 \\
0 & 0 & 0 & 0 & 0 & 0 & 1 \\
0 & 0 & 0 & 0 & 0 & 0 & 0 \\
0 & 0 & 0 & 0 & 0 & 0 & 0 \\
0 & 1 & 0 & 0 & 0 & 0 & 0 \\
0 & 0 & -1 & 0 & 0 & 0 & 0
\end{array}
\right)
\qquad
C_6 =\frac{1}{2}\left(
\begin{array}{ccccccc}
0 & 0 & 0 & 0 & 0 & 0 & 0 \\
0 & 0 & 0 & 0 & 1 & 0 & 0 \\
0 & 0 & 0 & -1 & 0 & 0 & 0 \\
0 & 0 & 1 & 0 & 0 & 0 & 0 \\
0 & -1 & 0 & 0 & 0 & 0 & 0 \\
0 & 0 & 0 & 0 & 0 & 0 & 0 \\
0 & 0 & 0 & 0 & 0 & 0 & 0
\end{array}
\right)
$$
$$
C_7 =\frac{1}{2}\left(
\begin{array}{ccccccc}
0 & 0 & 0 & 0 & 0 & 0 & 0 \\
0 & 0 & 0 & -1 & 0 & 0 & 0 \\
0 & 0 & 0 & 0 & -1 & 0 & 0 \\
0 & 1 & 0 & 0 & 0 & 0 & 0 \\
0 & 0 & 1 & 0 & 0 & 0 & 0 \\
0 & 0 & 0 & 0 & 0 & 0 & 0 \\
0 & 0 & 0 & 0 & 0 & 0 & 0
\end{array}
\right)
\qquad
C_8 =\frac 1{2\sqrt 3} \left(
\begin{array}{ccccccc}
0 & 0 & 0 & 0 & 0 & 0 & 0 \\
0 & 0 & -2 & 0 & 0 & 0 & 0 \\
0 & 2 & 0 & 0 & 0 & 0 & 0 \\
0 & 0 & 0 & 0 & 1 & 0 & 0 \\
0 & 0 & 0 & -1 & 0 & 0 & 0 \\
0 & 0 & 0 & 0 & 0 & 0 & -1 \\
0 & 0 & 0 & 0 & 0 & 1 & 0
\end{array}
\right)
$$
$$
C_9 =\frac 1{2\sqrt 3} \left(
\begin{array}{ccccccc}
0 & -2 & 0 & 0 & 0 & 0 & 0 \\
2 & 0 & 0 & 0 & 0 & 0 & 0 \\
0 & 0 & 0 & 0 & 0 & 0 & 0 \\
0 & 0 & 0 & 0 & 0 & 0 & 1 \\
0 & 0 & 0 & 0 & 0 & -1 & 0 \\
0 & 0 & 0 & 0 & 1 & 0 & 0 \\
0 & 0 & 0 & -1 & 0 & 0 & 0
\end{array}
\right)
\qquad
C_{10} =\frac 1{2\sqrt 3} \left(
\begin{array}{ccccccc}
0 & 0 & -2 & 0 & 0 & 0 & 0 \\
0 & 0 & 0 & 0 & 0 & 0 & 0 \\
2 & 0 & 0 & 0 & 0 & 0 & 0 \\
0 & 0 & 0 & 0 & 0 & -1 & 0 \\
0 & 0 & 0 & 0 & 0 & 0 & -1 \\
0 & 0 & 0 & 1 & 0 & 0 & 0 \\
0 & 0 & 0 & 0 & 1 & 0 & 0
\end{array}
\right)
$$
$$
 C_{11} = \frac 1{2\sqrt 3} \left(
\begin{array}{ccccccc}
0 & 0 & 0 & -2 & 0 & 0 & 0 \\
0 & 0 & 0 & 0 & 0 & 0 & -1 \\
0 & 0 & 0 & 0 & 0 & 1 & 0 \\
2 & 0 & 0 & 0 & 0 & 0 & 0 \\
0 & 0 & 0 & 0 & 0 & 0 & 0 \\
0 & 0 & -1 & 0 & 0 & 0 & 0 \\
0 & 1 & 0 & 0 & 0 & 0 & 0
\end{array}
\right)
\qquad
C_{12} = \frac 1{2\sqrt 3} \left(
\begin{array}{ccccccc}
0 & 0 & 0 & 0 & -2 & 0 & 0 \\
0 & 0 & 0 & 0 & 0 & 1 & 0 \\
0 & 0 & 0 & 0 & 0 & 0 & 1 \\
0 & 0 & 0 & 0 & 0 & 0 & 0 \\
2 & 0 & 0 & 0 & 0 & 0 & 0 \\
0 & -1 & 0 & 0 & 0 & 0 & 0 \\
0 & 0 & -1 & 0 & 0 & 0 & 0
\end{array}
\right)
$$
$$
C_{13} = \frac 1{2\sqrt 3} \left(
\begin{array}{ccccccc}
0 & 0 & 0 & 0 & 0 & -2 & 0 \\
0 & 0 & 0 & 0 & -1 & 0 & 0 \\
0 & 0 & 0 & -1 & 0 & 0 & 0 \\
0 & 0 & 1 & 0 & 0 & 0 & 0 \\
0 & 1 & 0 & 0 & 0 & 0 & 0 \\
2 & 0 & 0 & 0 & 0 & 0 & 0 \\
0 & 0 & 0 & 0 & 0 & 0 & 0
\end{array}
\right)
\qquad
C_{14} = \frac 1{2\sqrt 3} \left(
\begin{array}{ccccccc}
0 & 0 & 0 & 0 & 0 & 0 & -2 \\
0 & 0 & 0 & 1 & 0 & 0 & 0 \\
0 & 0 & 0 & 0 & -1 & 0 & 0 \\
0 & -1 & 0 & 0 & 0 & 0 & 0 \\
0 & 0 & 1 & 0 & 0 & 0 & 0 \\
0 & 0 & 0 & 0 & 0 & 0 & 0 \\
2 & 0 & 0 & 0 & 0 & 0 & 0
\end{array}
\right)
$$
\bibliographystyle{JHEP}
\bibliography {g2}
\end{document}